\documentclass[12pt]{article}
 \usepackage{amsmath,latexsym,amssymb,array,eufrak,graphicx,color}
 \newcommand{\1}[1]{{\normalsize\hfil #1}}
 
\title{HOW TO PLAY MACROSCOPIC QUANTUM GAME}
\date{}
\author{A.A.Grib\footnote{Department of Theoretical Physics and Astronomy,
A.I.Herzen State Pedagogical University,  Russia. e-mail:
Andrei-grib@mail.ru}, G.N.Parfionov\footnote{St.Petersburg State
University of Economics and Finances, St. Petersburg, Russia.
e-mail: GogaParf@gmail.com}}

\begin{document}
\maketitle
\begin{abstract}\footnotesize
Quantum games are usually considered as games with  strategies
defined not by the standard Kolmogorovian probabilistic measure
but by the probability amplitude used in quantum physics.  The
reason for the use of the probability amplitude or "quantum
probabilistic measure"\, is the nondistributive lattice occurring
in physical situations with quantum microparticles.   In our paper
we give examples of getting nondistributive orthomodular lattices
in some special macroscopic situations without use of quantum
microparticles.

Mathematical structure  of these examples is the same as that for
the spin one half quantum microparticle with two non-commuting
observables being measured.  So we consider the so called
Stern-Gerlach quantum games.  In quantum physics it corresponds to
the situation when two partners called Alice and Bob do
experiments with two beams of particles independently measuring
the spin projections of particles on two different directions In
case of  coincidences defined by the payoff matrix Bob pays Alice
some sum of money.  Alice and Bob can prepare particles in the
beam in certain independent states defined by the probability
amplitude so that probabilities of different outcomes are known.
Nash equilibrium for such a game can be defined and it is called
the quantum Nash equilibrium.

The same lattice occurs in the example of the firefly flying in a
box observed through two windows one at the bottom another at the
right hand side of the box with a line in the middle of each
window. This means that two such boxes with fireflies inside them
imitate two beams in the Stern-Gerlach quantum game.  However
there is a difference due to the fact that in microscopic case
Alice and Bob freely choose the representation of the lattice in
terms of non-commuting projectors in some Hilbert space.  In our
macroscopic imitation  there is a problem of the choice of this
representation(of the angles between projections). The problem is
solved by us for some special forms of the payoff matrix.  We
prove the theorem that quantum Nash equilibrium occurs only for
the special representation of the lattice defined by the payoff
matrix.  This makes possible imitation of the microscopic quantum
game in macroscopic situations.  Other macroscopic situations
based on the so called opportunistic behavior leading to the same
lattice are considered.
\end{abstract}

In this paper we continue the investigation \cite{1} of
macroscopic situations described by the same  mathematical
formalism as some simple (spin one half and spin one) quantum
systems.  In these situations stochasticity is described by some
wave function as vector in finite dimensional Hilbert space  with
non-commuting operators in it as observables.

So one has the complementarity property for such systems.  These
situations can occur in economics \cite{2,3}, biology \cite{4} etc
so that chance in these sciences must not necessarily be described
by the standard Kolmogorovian probability measure as it is usually
supposed to be but by the more general quantum formalism.  In
\cite{1} it was shown that new type of Nash equilibrium can arise
in these cases.  Differently from the microworld the Planck's
constant does not play any role in our examples.

\section{\hspace{-20pt}. The Firefly in a Box}
Here  we consider some other than in  our papers \cite{1,5,6}
example named ''the firefly in a box'' \cite{7}.  This example
will be used by us in order to show the connection between the
Boolean lattice with probabilistic description of chance and the
non Boolean nondistributive lattice with the quantum probability
measure on it. The example is the simplification of the well known
more complex Foulis ''firefly in a box'' case \cite{8,9}.  We take
it because in our paper \cite{1} we found  Nash equilibrium for a
quantum game based on Hasse diagram for this simplified case. It
occurred that in quantum case there are Nash equilibrium more
profitable than the classical ones.  These Nash equilibrant
correspond to realizations of the nondistributive lattice in terms
of certain noncommutative projectors in Hilbert space of the spin
one half system. We shall not discuss other known examples (Wright
urn \cite{10}) leading to nondistributive lattices because we did
not investigate Nash equilibrium for these examples. The rule for
the quantum macroscopic game for our example can be formulated as
follows.
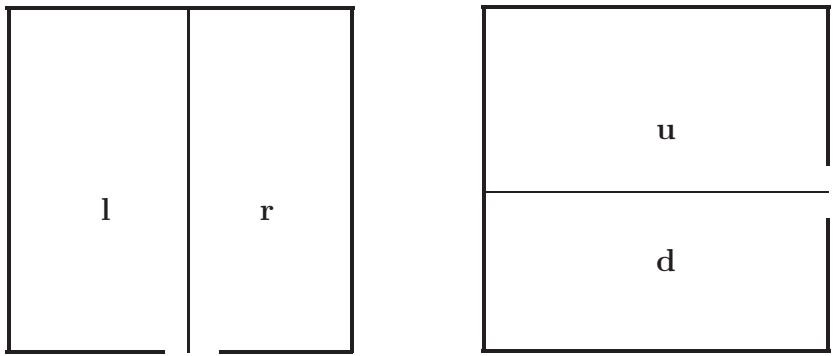
\begin{figure}[h]
\begin{center}
 \begin{picture}(300,130)
 \put(53,0){\line(0,1){130}}
 \linethickness{1pt}
 \put(80,50){\bf r} \put(20,50){\bf l}

 \put(-16,130){\line(1,0){132}}

 \put(-16,0){\line(1,0){60}}
 \put(65,0){\line(1,0){50}}

 \put(-15,0){\line(0,1){130}}
 \put(115,0){\line(0,1){130}}
 \end{picture}

 \begin{picture}(-60,-50)
 \put(-15,75){\line(1,0){130}}
 \linethickness{1pt}
 \put(50,45){\bf d} \put(50,95){\bf u}

 \put(-16,145){\line(1,0){132}}
 \put(-16,15){\line(1,0){132}}

 \put(-15,15){\line(0,1){130}}
 \put(115,15){\line(0,1){50}}
  \put(115,85){\line(0,1){60}}
 \end{picture}
\end{center}
\vspace{-25pt}\caption{The firefly in  a box with two windows.}
\label{fig1}
\end{figure}
A firefly (surely a macroscopic agent) is roaming around   a box
and some observer (other macroscopic agent) can see it either
through the  window at the bottom of the box or through the window
at the right side of it.  Each window has a thin line
perpendicular to it drawn at it's center so that an observer can
see the firefly in one or another halves of the box.

An observer cannot look at the same moment at two windows  at once
so that one has two incompatible experiments.  Let us call
possible observable situations as "left", "right", "up" and
"down".  The outcomes of the experiments can be described by the
nondistributive lattice employed by Birkhoff and von Neumann
\cite{11} for the spin one half system with two complementary
observables -- (different spin projections) -- being measured.
This is an orthocomplemented  lattice.  All rigorous mathematical
definitions can be found in \cite{12}. However for better
understanding of the paper we must give some necessary definitions
and conjectures here.

The lattice L is a partially ordered set (S, $\leq$) with two
operations $\vee, \wedge$ so that each pair $x, y \in L, x \neq y$
has a supremum $x\vee y$ and an infimum $x \wedge y$. There are
elements $\oslash \in L, I \in L$ such that $x \vee \oslash = x$,
$x \wedge \oslash$, $x \vee I = I$, $x \wedge I=x$.

The lattice is complemented if for $\forall x \in L$ exists at
least one complement $x'$ such that $x \wedge x' = \oslash$, $x
\vee x' = I$.  The elements of the lattice are orthogonal $x \bot
y$ if $x \leq y'$.  The operations $\wedge, \vee$ can sometimes be
understood as logical disjunction and conjunction. Then it is
supposed that if x is true than $x \vee y$ is true, if $y$ is true
then $x \vee y$ is true.  However in quantum logic ''if'' is not
equal to ''always if'' (iff).  So in general it is not correct to
think that from $x \vee y$ true follows that either of them is
true.

Negation in quantum logic is realized through orthogonality. There
is some discussion in literature on the problem of logical
interpretation of lattices (see \cite{13}) for the general case so
that different views arise from different definitions.  For
Boolean sublattices of the non Boolean lattice it is possible to
give usual logical interpretation of the lattice operations.

The success of quantum physics shows that the idea of Birkhoff and
von Neumann of the nondistributive lattice with quantum
probabilistic measure on it as justification of use of the
probability amplitude is a correct one. So it seems reasonable to
conclude that in other cases when the same lattices occur one must
obtain the quantum mechanical mathematical formalism.  One can
draw the so called Hasse diagram for the lattice so that lines
correspond to partial order,going up one can obtain intersection
of lines at $\vee$, going down one can obtain intersection
at~$\wedge$.
\begin{figure}[ht]
\begin{center}
\begin{picture}(120,120)
\put(45,94){\circle*{4}} \put(43,105){I} \put(0,50){\circle*{3}}
\put(-10,55){$l$} \put(30,50){\circle*{3}} \put(25,55){$r$}
\put(60,50){\circle*{3}} \put(65,55){$u$} \put(90,50){\circle*{3}}
\put(95,55){$d$} \put(45,6){\circle*{4}}
\put(42,-10){$\varnothing$} \put(0,50){\line(1,1){45}}
\put(0,50){\line(1,-1){45}} \put(90,50){\line(-1,1){45}}
\put(90,50){\line(-1,-1){45}} \put(30,50){\line(1,3){15}}
\put(30,50){\line(1,-3){15}} \put(60,50){\line(-1,3){15}}
\put(60,50){\line(-1,-3){15}}
\end{picture}
\end{center}
\vspace{-5pt}\caption{Hasse diagram of the nondistributive lattice
of the firefly in a box example. } \label{fig3}
\end{figure}
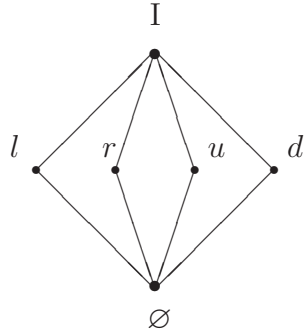
Here "l", "r", "u", "d"\, are elements (logical atoms) of the
lattice describing different experimentally testable propositions
for the firefly on Fig.1.  Elements "l" and "r" as well as "u" and
"d" are orthogonal. One also has for the two lattice operations
$\wedge$ ("and"), $\vee$ ("or")
$$
l\vee r = u \vee d = l \vee u = r \vee d  = l \vee d  = r \vee u = I
$$
$$
l \wedge r = l \wedge u = l \wedge d = r
 \wedge u = u \wedge d = r \wedge d =  \oslash
$$
which means that "l" or "r" is always true  while "l" and "r" is
always false etc.  For example "r" and "u" is always false because
"experimentally" there is no such observable element  at the
disposal of the observer due to the impossibility of simultaneous
observation of the corresponding situations.  The lattice is
nondistributive because
$$
l \wedge (r \vee d)=l\wedge I = l \neq
 (l\wedge r)\vee (l\wedge d) = \oslash \vee \oslash = \oslash .
$$
If the firefly randomly moves inside the box the observer can
describe the outcomes of his observations  as some representation
of the nondistributive lattice in terms of projectors ("yes - no"
questions) in two dimensional real space.  He (she) defines the
"quantum" probability of the outcomes from some wave function $ |
\Psi \rangle $  taken as the two dimensional vector by
\begin{equation}
w_{\alpha}= \langle \Psi | \hat{P_{\alpha}}
 | \Psi \rangle , \alpha \in \{ l, r, u, d \} ,
\end{equation}
here one has
\begin{equation}
w_l + w_r = w_u + w_d = 1.
\end{equation}
So one has the wave function and non-commuting operators -- the
projectors $\hat{P_{\alpha}}$ for the ''firefly in a box''
example.   To organize the game first consider the game in which
quantum  microparticles are used.  Call it the Stern-Gerlach
quantum game \cite{1}.  Two partners Alice and Bob are sitting
close to accelerators and prepare two beams of particles (protons)
with spin one half.  Then they do the Stern-Gerlach experiment
measuring two different spin projections of their particles. There
is some payoff matrix given by the table \ref{payoff matrix}.
\begin{table}[h]\large
\begin{center}\caption{\footnotesize payoff of Alice}\label{payoff matrix}
\begin{tabular}{|p{15mm}|p{45.1mm}|}
\hline
\1{\scriptsize strategies}&\1{\small Bob}\\
\end{tabular}\vspace{-1mm}
\begin{tabular}{|p{15mm}|p{8mm}|p{8mm}|p{8mm}|p{8mm}|}
\hline\1{\small Alice}&
\1{\textbf{1}}&\1{\textbf{2}}&\1{\textbf{3}}&\1{\textbf{4}}\\
\hline

\1{\textbf{1}}&\1{0}&\1{0}&\1{$c_3$}&\1{0}\\ \hline

\1{\textbf{2}}&\1{0}&\1{0}&\1{0}&\1{$c_4$}\\ \hline

\1{\textbf{3}}&\1{$c_1$}&\1{0}&\1{0}&\1{0}\\ \hline

\1{\textbf{4}}&\1{0}&\1{$c_2$}&\1{0}&\1{0}\\ \hline
\end{tabular}\end{center}\vspace{-15pt}
\end{table}
The meaning of it is that in case Alice gets result \(\textbf{1}\)
and  Bob gets \(\textbf{3}\) Bob pays to Alice the sum of money
$c_3$ prescribed by the payoff matrix etc. There are some
frequencies \(p_1\) of getting \(\textbf{1}\) by Alice in a series
of her measurements and there are frequencies \(q_3\) of getting
\(\textbf{3}\) by Bob. These frequencies are defined by the
probabilities of certain outcomes which can be calculated by the
rules of quantum physics if one knows the wave functions of
particles prepared by Alice and Bob in their beams. The average
profit of Alice can be calculated as
\[\overline{H}_A=c_1p_3q_1+c_3p_1q_3+c_2p_4q_2+c_4p_2q_4
\]
\begin{equation}\label{for_prob}
    p_1+p_3=1,\quad p_2+p_4=1,\quad q_1+q_3=1,\quad q_2+q_4=1
\end{equation}
The strategies of Alice and Bob in this game are described by the
wave functions of particles in their beams. It is supposed that
Alice and Bob have special experimental setups to produce their
particles in certain states with some fixed wave function. However
they are not free in their choice of measuring spin projections.
It is supposed that both partners know what projections are to be
measured, For macroscopic situations described by the same lattice
as the Stern-Gerlach quantum game type one can be interested to
find answers on the questions: what is the meaning of the
"preparation"\, of the wave function and what is the meaning of
measuring different non-commuting observables from the point of
view of our macroscopic agents?

To find the  answer one must take into account the fact that this
lattice can be embedded into some Boolean lattice.  The physical
meaning of the embedment is simple: the observer cannot check some
situations described by some more general Boolean lattice due to
the character of  his (her) experiments.  These elements of the
lattice are "hidden variables"\, for the observer and the
non-distributivity of the lattice is the payment for his (her)
''ignorance''.  A general theory of realizing of quantum logical
lattices as ''concrete logics'' obtained from Boolean lattices was
developed in \cite{15}.

Surely  due to Kochen-Specker's theorem \cite{15} not all quantum
logical lattices can be embedded into Boolean lattices, that is
why quantum theory of microparticles is basic and is not a hidden
variable theory.  The other objection is breaking of Bell's
inequalities for relativistic systems if entangled states are
considered \cite{16}.

However for our  aim to find macroscopic situations with
macroscopic agents with  behavior described by the quantum rules
(i.e.  by the Born-Luders rule for calculation of probabilities)
the class of such ''quasi-classical'' lattices \cite{5,7} embedded
in Boolean ones is wide enough.  To construct the  Boolean lattice
divide the box on four parts.
\begin{figure}[ht]
\begin{center}
\begin{picture}(100,130)
\linethickness{1pt}

\put(15,95){$w_2$}

\put(15,30){$w_1$}

\put(75,95){$w_4$}

\put(75,30){$w_3$}

\put(-16,65){\line(1,0){132}} \put(50,0){\line(0,1){130}}

\put(-16,130){\line(1,0){132}}

\put(-16,0){\line(1,0){132}}

\put(-15,0){\line(0,1){130}}

\put(115,0){\line(0,1){130}}
\end{picture}
\end{center}
\vspace{-5pt}\caption{Probability of different firefly in a box
positions from it's ''own'' point of view. }\label{fig1}
\end{figure}

The firefly can occur in  any of the four parts.  Construct the
Boolean lattice based on elements \(1, 2, 3, 4\) as it's  atoms.
One  obtains the   nondistributive lattice of the Fig.2 from the
distributive lattice by considering only composite elements "l",
"r", "u", "d"\, of the second row of the boolean lattice while the
atomic elements \(1, 2, 3, 4\) as well as the third row composed
from triples are unobservable.

So the main lesson is that one can  expect obtaining of the
quantum rules in situations when one has stochastic processes
which are secondary to some basic non observable ones.  This is
typical for situations on the stock market, in some complex
biological systems etc.  The other important feature is
complementarity as impossibility of checking all properties at the
same moment of time. What is the connection of the  probabilities
on the Boolean lattice and the quantum probability on the
nondistributive one?

Denoting $w_a:\;\; w_1, w_2, w_3, w_4$ the probabilities  for
atoms of the Boolean lattice one obtains the equations
\begin{equation}
\begin{array}{llll}
{\displaystyle w_1+w_2=w_l}\\
{\displaystyle w_3+w_4=1-w_l}\\
{\displaystyle w_1+w_3=w_d}\\
{\displaystyle w_2+w_4=1-w_d}
\end{array}
\end{equation}
From these equations it is easy  to see that the "strange"\,
quantum rule for one and the same object (the  firefly) is not
strange at all if it  is considered not for the elementary events
but for the complex ones!
\begin{equation}
w_l+w_r+w_u+w_d=2
\end{equation}
It seems that any distribution  $w_1, w_2, w_3, w_4$  leads to
some fixed wave function and some fixed representation of the
lattice in terms of projectors on some fixed directions on the
plane.  However one can see that distributions leading to
appearance of two ones or zeros for complex events of the second
floor are prohibited due to the definition of disjunction in the
quantum logical lattice while it is possible in the Boolean case.
This means that logical atoms of the first floor cannot be
definitely determined from the point of   the  observer for the
quantum logical case being totally "hidden"\, for him (her).  One
can ask the question: what prohibits the firefly to occur in the
corner \textbf{1}?

Is it possible for the observer  to get \textbf{1} for the outcome
"left"\, and \textbf{1} for the outcome "down"? The answer is
surely positive.  But for the quantum system it is impossible to
get eigenfunction for non-commuting spin operators.  Is it a
contradiction?

The answer is "no"! It  is here where  something like the wave
packet collapse idea comes into the play.  Quantum theory does not
forbid to  get in observations of complementary observables
positive results.  It only says that if one prepares the
wavefunction as an "eigenfunction of the projector on the "left"
part and performs many observations of complementary observables
then one obtains some  probability distribution for the
complementary observable ''up''-''down''.  This distribution is
not 1, 0 because the firefly has some freedom and the only
instruction for him given on the preparation stage is to be
"somewhere" in the "left" part.  The observer cannot always let
him be in the corner1,so in many experiments with the same "left"
instruction the results will be sometimes "up", sometimes "down"
with frequencies obtained from the wave function and the
representation of observables in the form of operators.

If the distributions $w_l, w_d$  are known the distribution $w_1,
w_2, w_3, w_4$ obtained from it is not unique: it is defined up to
some arbitrary $w_i$ where $w_i$ is some probability for fixed
$i$.

This is just the manifestation  of "indefiniteness"\, of the
quantum situation in comparison to Boolean one as we just said
before.

\section{\hspace{-20pt}. Stern-Gerlach quantum game}
The macroscopic quantum game considered  previously in \cite{1}
called the macroscopic Stern-Gerlach quantum game can be organized
for the "firefly in a box"\, case as follows.

There are two partners Alice and Bob  and two boxes with fireflies
there.  Alice can try to choose some classical probability
distribution for Boolean elementary outcomes 1, 2, 3, 4 with
limitations discussed before.  This can be done by "training"\, of
the firefly stimulating it to come more often to this or that part
of the box.  For example  any  observation by Alice of the part of
the box is accompanied by the flash of light with some
prolongation in time stimulating the firefly to react.  Different
times of observation result in different frequencies for the
firefly to occur in some part of the box.   Supposing that Alice
is interested only in  $l, r, u, d$ outcomes she cannot define
exact distributions for Boolean elementary outcomes but only some
class of it.  However from the quantum point of view  definition
of frequencies for $l, r, u, d$ means definition of the wave
function and the representation of the  atoms of the non Boolean
lattice in terms of non-commuting projectors, i. e.  definition of
the angle between different spin projections.  This can be called
"the preparation stage".

In special cases getting \(1,0\)  for some of the complementary
possibilities she can speak about some fixed wave function as an
eigenfunction of some spin projection operator. However in general
case it is not the case.

Same manipulations are made by  Bob.  However neither Alice nor
Bob have knowledge of the training procedures of the partner.

Then the game  begins.  Alice and Bob with  their trained
fireflies look at the results of their observations accompanied by
flashes of light.  In cases  defined by the rules of the game when
for example Alice gets some fixed result ''$\alpha$'' while Bob
gets ''$\beta$'' Bob must pay money to Alice etc.

There is some payoff matrix defining in what  cases Bob pays Alice
some money as it was defined in \cite{1}.  The  profit depends on
the frequencies of the outcomes.  The average profit is calculated
by using  the quantum rule for projectors $\hat{P_{\alpha}^a}$ for
Alice, $\hat{P_{\alpha}^b}$ for Bob
\begin{equation}
\begin{array}{c}
\displaystyle{\overline{H}=
\langle \Psi_A | \langle \Psi_B | \hat{H} | \Psi_B \rangle \Psi_A \rangle; } \\
\displaystyle{\hat{H}= c_3 \hat{P_u^a} \otimes \hat{P_d^b} + c_1
\hat{P_d^a} \otimes \hat{P_u^b} + c_4 \hat{P_l^a} \otimes
\hat{P_r^b} + c_2 \hat{P_r^b} \otimes \hat{P_l^b}.  }
\end{array}
\end{equation}
which in terms of the "Boolean philosophy" means calculation of
\begin{equation}\label{for_6}
\overline{H}= c_3 w_u w_d^b + c_1 w_d w_u^b + c_4 w_l w_r^b + c_2 w_r w_l^b,
\end{equation}
\[w_u+ w_d=w_l+ w_r=w_u^b+ w_d^b=w_l^b+ w_r^b=1
\]
where $w_{\alpha}, w_{\alpha}^b$ are probabilities used by Alice
and Bob. One can recognize in~\eqref{for_6} formula
\eqref{for_prob} with $w_\alpha^b$ playing the role of $q_\alpha,
p_\alpha$.

Here one must make some remarks about  some peculiar features of
this calculation. The representation of the non Boolean lattice in
terms  of non-commuting projectors is not unique.  It is defined
up to some angles~$\theta, \tau$.  Existence of different
representations of the non Boolean lattice parameterized by the
angles is the manifestation of the freedom of the observer to
choose measurement of any spin observable for the real quantum
system.  There is also another freedom for the observer manifested
in the choice of the wave function.

Let us defined
\begin{equation*}
\begin{array}{c}
  w_u = \cos^2\alpha,\qquad w_l = \cos^2(\alpha - \theta) \\
  w_u^b = \cos^2\beta,\qquad    w_r^b = \cos^2(\beta - \tau)
\end{array}
\end{equation*}
If after  this definition of the angle Alice and Bob cannot change
the angles then their freedom is now limited by choosing  only
some special distributions for the firefly  in the complementary
experiment.  The "quantum logic"\, leads to  arising of a special
"quantum correlation"\, between complementary observations.  This
correlation according to \cite{1} can be expressed by the
constraint
\begin{equation}
{\frac{(w_u+w_l-1)^2}{cos^2\theta} +
\frac{(w_l-w_u)^2}{sin^2\theta}} = 1.
\end{equation}
Different choice of  the angles leads to the different constraint.
Different Nash equilibrium can be found for different angles.  It
was shown in \cite{1} for the Hasse diagram considered in this
paper that some  of these equilibrium are more profitable for
partners others are less. So one can put the hypothesis that it is
the more profitable equilibrium that can play the role of the
principle of choice of the representation of the lattice.

\bigskip

\section{\hspace{-20pt}. Eigenequilibrium}
There is some special  case of the payoff matrix discovered in our
paper~\cite{1} in which our macroscopic quantum game is totally
defined. This case was called by us the case of
''eigenequilibrium''.

In quantum game on the quadrangle \cite{1} with payoff matrix (see
tabl.\ref{payoff matrix}) the average payoff of Alice can be
written as
\[
\langle H\rangle=\frac{1}{4}\,(g(x,y)+\mathrm{tr}\, C)
\]
where
\begin{equation*}\label{payh}
    g(x,y)=-\langle x,A y\rangle+\langle
x,M^\dag_\theta \omega \rangle-\langle M^\dag_\tau \omega, y
\rangle,
\end{equation*}
$A=M^\dag_\theta CM_\tau$, and $x$, $y$ unit vectors on the plane:
$|x|=1,\; |y|=1$.
 Here
\[
M_\varphi=\left[%
\begin{array}{cr}
  \cos\varphi & -\sin\varphi\\
  \cos\varphi & \sin\varphi \\
\end{array}%
\right],\quad n=c_{1}+c_{3},\; m=c_{2}+c_{4}\]
\[
\omega=\left[%
\begin{array}{c}
  c_3-c_1 \\
  c_4-c_2 \\
\end{array}%
\right],
\qquad C=\left[%
\begin{array}{cc}
  n & 0 \\
  0 & m \\
\end{array}%
\right]
\]
An equilibrium \((x,y)\) is called \textit{eigenequilibrium}, if
it is an eigenvector of the matrix
\[
\mathcal{A}=\left[\!\!%
\begin{array}{cc}
  O      & A \\
  A^\dag & O \\
\end{array}%
\!\!\right]
\]
The following proposition proved in \cite{19}.

\textbf{Proposition 1.} {\it If the eigenequilibrium exists, then
$\omega$ is a common eigenvector of the matrices $CM_\theta
M^\dag_\theta$ ¨ $CM_\tau M^\dag_\tau$. }

A game is said to be {\it non-degenerate}, if
\begin{equation*}\label{Delta}
    \Delta=\left|%
\begin{array}{cc}
  n & m \\
  \omega_1^2 & \omega_2^2 \\
\end{array}%
\right|\neq 0
\end{equation*}

\textbf{Proposition 2.} {\it If the game is non-degenerate, then
the necessary condition for the eigenequilibrium to exist is the
coincidence of the angular parameters \hbox{$\theta=\tau$}. In
this case their values are completely determined by the payoff
coefficients of the game $\{c_j\}$:
\begin{equation}\label{theta-tau}
\cos2\theta=\cos2\tau=\frac{(m-n)\omega_1\omega_2}{\Delta}
\end{equation}}
Further finding \textit{eigenequilibrium} of
\textit{non-degenerate} games, calculate $\theta$ using
\eqref{theta-tau} and put $M=M_\theta$, \(z=M^\dag\omega\). In
this case $A=A^\dag=M^\dag CM$ and the matrix $A$ non-negatively
defined.

\textbf{Proposition 3.} {\it \textsc{(Existence theorem)} Let a
vector \(\omega\) be an eigenvector of the matrix \(CMM^\dag\) and
\(\langle Az, z\rangle\leqslant|z|^3 \). Then the strategies
\(x=y=z/|z|\) form an eigenequilibrium.}

\textbf{Proposition 4.} {\it \textsc{(Multiple Nash-equilibrium)}
Let a vector \(\omega\) be an eigenvector of the matrix
\(CMM^\dag\) and \(\langle Az, z\rangle=|z|^3 \). Then there are
two eigenequilibrium \hbox{\(x=y=z/|z|\)} ¨ \(x=-z/|z|,\;
y=z/|z|\). }

\textbf{Proposition 5.} {\it \textsc{(Uniqueness theorem)} Let
there is a game with a non-degenerate equilibrium \(\langle Az,
z\rangle\neq |z|^3 \). Then all possible equilibrium are exhausted
by it. }

So, it occurs that optimal strategies of the players are defined
not so by the representation of the ortholattice as by the
ortholattice itself and by the payoff structure of the game.

For this case as in quantum Stern-Gerlach quantum game the angles
are prescribed by the rule of the game and the only choices for
Alice and Bob training their fireflies concern probability
distributions satisfying the constraint with this angle.

The optimal choice corresponds to Nash equilibrium existing for
this angle.  For other choice of the angle Nash equilibrium does
not exist and clever Alice and Bob will not use them at all.

\section{\hspace{-20pt}. Example of the multiple\\ quantum Nash-equilibrium}
For \[c_1=1;\quad c_2=2;\quad c_3=99;\quad c_4=98
\] the angles is
equal $\theta=\tau=45^{\circ}$ and the optimal strategies of Alice
and Bob are \[p_1=q_1\thickapprox 0,857;\quad p_2=q_2\thickapprox
0,622;\qquad \langle H\rangle=50
\]

\noindent For \[c_1=1;\; c_2=2;\; c_3=9;\; c_4=8
\] the angles is
equal  $\theta=\tau=45^{\circ}$ and there are \textit{two}
eigenequilibrium.

First equilibrium:
\[p_1=q_1=0,9;\quad p_2=q_2=0,8;\quad \langle
H\rangle=5
\]

Second equilibrium:
\[ p_1=0,1;\; q_1=0,9;\quad p_2=0,2;\; q_2=0,8;\quad \langle
H\rangle=5
\]

\section{\hspace{-20pt}. Quantum equilibrium\\ against the classical one}
If one considers the classical game with the same payoff matrix
one can obtain~\cite{1} for the average profit:
$$
h=(c_{1}^{-1}+c_{3}^{-1})^{-1}+(c_{2}^{-1}+c_{4}^{-1})^{-1}
$$
It can be shown in some cases that the quantum equilibrium is more
profitable than the classical one. Consider the following example:
$c_{1}=1$, $c_{2}=9$, $c_{3}=10$, $c_{4}=2$ the optimal strategies
of Alice and Bob are
\[p_1=q_1=\frac{130+9\sqrt{130}}{260}\thickapprox 0.895, \quad
p_2=q_2=\frac{130-7\sqrt{130}}{260}\thickapprox 0.193
\]
The optimal profit in the classical game is smaller than in the
quantum one: $\langle h\rangle=28/11,\; \langle H\rangle=11/4$.

\paragraph{Acknowledgments.}
The authors are  indebted to the  Ministry of Science and
Education of Russia, grant RNP 2.1.1.6826 for financial support of
the work on this paper.

\end{document}